\documentclass[twocolumn,showpacs,showkeys,a4paper,10pt,final,nofootinbib,prb]{revtex4}

\usepackage{graphicx}
\usepackage{subfigure}
\usepackage{dcolumn}
\usepackage{tabularx}
\usepackage{braket}
\usepackage{bm}
\usepackage{amsmath,amssymb,amsfonts,mathrsfs}
\usepackage{wrapfig,epsfig}
\usepackage{tikz}
\usepackage{pgf}

\newcommand{\de}{\mathrm{d}}  
\newcommand{\bv}{\mathbf}  

\begin{document}

\title{Local free energy approximations for the coarse-graining of adsorption phenomena}
\author{Federico G. Pazzona}%
\email{fpazzona@uniss.it}%
\author{Giovanni Pireddu}%
\author{Andrea Gabrieli}%
\author{Alberto M. Pintus}%
\author{Pierfranco Demontis}%
\affiliation{Dipartimento di Chimica e Farmacia, Universit\`a degli Studi di Sassari, via Vienna 2, 01700 Sassari, Italy}


\begin{abstract}
  We investigate the coarse-graining
  of host-guest systems under the perspective of the 
  local distribution of pore occupancies, 
  along with the physical meaning and actual 
  computability of the coarse-interaction terms.
  We show that the widely accepted approach, in which 
  the contributions to the free energy given by the 
  molecules located in two neighboring pores
  are estimated through Monte Carlo simulations where 
  the two pores are kept separated
  from the rest of the system, leads to inaccurate results 
  at high sorbate densities.
  In the coarse-graining strategy that we propose,
  which is based on the Bethe-Peierls approximation,
  density-independent interaction terms are instead 
  computed according
  to local effective potentials that take into account the 
  correlations between the pore pair and its surroundings
  by means of mean-field correction terms, without the need of
  simulating the pore pair separately.
  Use of the interaction parameters obtained this way
  allows the coarse-grained system to reproduce more closely the
  equilibrium properties of the original one.
  Results are shown for lattice-gases where the local
  free energy can be computed exactly, and for
  a system of Lennard-Jones particles under the effect of
  a static confining field.
\end{abstract}



\maketitle

\section{Introduction \label{sec:Intro}}

Despite the increasing availability of computing power,
molecular simulations with atomistic detail suffer from severe 
limitations in the length and time scales, even when
the interaction field is classical.

To reduce the number of degrees of freedom involved in 
a simulation, thus allowing simulations to be carried 
out over wider scales, is the scope of coarse-graining
techniques.
In the coarse-graining of a molecular system, the original, 
fine-grained (FG) interaction field is mapped into an 
effective field that depends on a smaller number of variables,
and the mapping is carried out in such a way that some selected
properties of the FG system and of the coarse-grained (CG) model 
reasonably match.
Since such properties are defined on a scale that is usually larger 
than the one at which the FG system evolves,
this comes at the cost of a certain loss of information.

In the literature, the coarse-graining of molecular systems 
is approached in a variety of ways.
Many of such approaches are \emph{topological}, that is, each 
CG coordinate groups together 
several atoms of the FG system, and interacts with the other
CG coordinates through effective fields that can be built from 
structure,~\cite{Soper1996,Shell2008,Bilionis2014} or \emph{via} a force-matching 
procedure~\cite{Izvekov2005,Izvekov2005b,Noid2008,Das2009} 
(the two approaches leading to the same results~\cite{Rudzinski2011}),
or through iterative Boltzmann inversion,~\cite{Reith2003,Hadley2010,Moore2014}
or else through Gaussian Approximation Potentials,~\cite{John2017}
and cluster expansion techniques,~\cite{Tsourtis2017} just to 
mention some---we do not mean to make an exhaustive list here.
Besides topological strategies,
a \emph{spatial} coarse-graining approach also exists, 
which maps portions of a continuous simulation space,
as well as groups of FG discrete \emph{sites}, into a coarser
lattice of \emph{cells}.~\cite{Ma1976,HoMM,Katsoulakis2003,Chatterjee2004,Dai2008,Liu2012,Katsoulakis2014,delaTorre2011,Israeli2006}
A \emph{cell state} can be constructed out of what it contains, e.g.,
for a molecular systems, that could be, very naturally, the number of 
molecular centers-of-mass of each chemical species that occupy its 
physical space.

It is the application of the latter spatial approach to the 
coarse-graining of adsorption phenomena at equilibrium
that we intend to discuss in this work.
By keeping in mind the picture of small \emph{guest} molecules adsorbed 
inside the pores of some \emph{host} material, we will identify each 
cell as a pore, and the state of each one of them as the 
\emph{occupancy}, which we define as 
the number of molecular centers-of-mass it hosts---not to be confused with the
\emph{loading}, with which we will indicate the average pore occupancy.
For simplicity, we will discuss the case of only one guest chemical species
in the system, but extension to multispecies models is straightforward.

Occupancy-based models of adsorption/diffusion, where
a CG interaction field is defined over local occupancies
in the nearness of discrete locations, rather than on 
fine-grained atomistic configurations,
are frequently encountered in the literature on host-guest 
systems.~\cite{vanTassel1993,Snurr1994,Auerbach1998,Czaplewski1999,TuncaFord1999,TuncaFord2002,Tunca2003,Tunca2004}
According to how detailed should the CG model be,
these locations may represent \emph{adsorption sites},
that usually can be empty or occupied by one guest, or pores,
that often can be occupied by more than one guest.
Depending on the affinity between the host material and the 
guest species,
adsorption sites may emerge naturally within the adsorption 
pores as well-defined locations, that bind the guest 
molecules more strongly than others.
This is the case for, e.g., benzene in silicalite~\cite{Snurr1994},
methane in the zeolite ITQ-29 (a.k.a.~ZK4)~\cite{Demontis1997_JCPB},
and benzene in zeolite Na-Y,~\cite{Auerbach1998,Auerbach2000,Auerbach2004}
just to mention some.
In such cases, a CG version of the grand canonical partition 
function can be constructed
by modeling the adsorption sites as mutually exclusive lattice nodes
equipped with a proper adsorption energy, while the guest-guest 
interactions can be represented as pairwise-additive free 
energies (such assumption provides a satisfactory approximation
especially at low densities, where many-body contributions are 
proved to be relatively unimportant~\cite{Das2009}), plus, 
if necessary to improve the model quality, inclusion of next 
neighbor interaction terms.~\cite{Czaplewski1999}
Further additional interactions, expressed in the form
of dependency on some collective (but still \emph{local}) variables,~\cite{Wagner2017}
may be also necessary.
In any case, it is preferable to work with
\emph{local}, rather than global interaction energies,
because, besides a number of other general 
drawbacks~\cite{Louis2002},
the dependence of effective potentials on global density 
imposes severe limitations to transferability, e.g.~to
inhomogeneous systems.~\cite{DeLyser2017}

Identifying the \emph{pores} of an ordered microporous material,
rather than adsorption sites, as the elementary units of a 
discrete space domain, represents an even coarser description of adsorption.
A pore is usually allowed to contain more than
one guest molecule, and this makes the resulting CG model a
so-called `multiparticle lattice-gas'.~\cite{ChopardDroz}
When strong confinement holds and the density is not high, 
the correlation between molecules located inside different pores 
is often found to be weaker than inside the same pore.
If that is the case, a CG interaction field can be 
satisfactorily formulated as a function of individual, 
uncorrelated pore occupancies, at least at room temperature
(depending on the system, this might happen to be not true 
at lower temperatures).~\cite{Pazzona2014}
Assuming such a strict locality of interactions allows 
for a very simple and efficient description of both the 
thermodynamics and the kinetics of particle pore-to-pore 
jumps.~\cite{Pazzona2011,Becker2013}
If accounting for pore-pore interactions becomes necessary,
pairwise additivity can
still be assumed at low densities, so that we can
factorize the resulting CG grand partition
function into elementary terms that, in principle, can be estimated out of
a proper statistical sampling of the FG system 
itself.

When dealing with the calculation of approximated partition 
functions in general,~\cite{Vieth1995}
factorization is really a crucial point.
Somewhat radical, oversimplifying approximations usually lead to `friendly'
CG partition functions, made of independent 
(or nearly independent) factors that often can
be evaluated easily, but often
such approximations suffer from a narrow range 
of applicability.
On the other hand, more broadly acceptable 
approximations usually go along with a much more 
difficult evaluation of the constituting 
factors of the CG partition function---ironically, estimating them
might end up requiring the use of \emph{further} 
approximations.

Therefore, a balance needs to be found between the 
accuracy of the approximations on which the CG model 
is based, and the actual computability of its 
parameters.
In the present paper, we discuss
the formulation of a CG grand partition function for 
host-guest systems in which effective interactions, which
are portrayed by both self- and pair-interaction terms, 
are defined over pore occupancies.
We propose a modification of an existing CG model~\cite{Tunca2003}
of interactions of such kind, that significantly 
widens its applicability to a larger density range.
In our formulation, effective pair interactions are, 
although still local, related to the occupancy correlations 
that can be observed between neighboring pores 
within a given range of densities.

Our discussion will proceed as follows.
First, in Section~\ref{sec:CGpf} we will briefly resume 
how the CG grand partition function is formulated,
based on pore occupancies rather than molecular positions.
In Section~\ref{sec:IPA} we will formulate a relation between 
local CG interactions and occupancy distributions
in the FG system, with mean-field corrections taking into account 
the effect of the neighborhood of any single pore and of any pore pair.
In Section~\ref{sec:compare} we will compare our basic CG relations
to an earlier, simpler theory were the surroundings of a pore
pair is not taken account of in any way, and we will
also show how, under less general circumstances, 
the parameterization we propose here reduces to the model
we proposed in a previous work.~\cite{Pazzona2014}
In Sections~\ref{sec:lattice} and~\ref{sec:LJ} we will apply our
method to the coarse-graining of FG systems of two kinds:
a lattice-gas where local free energies can be computed exactly, 
and a Lennard-Jones system of united-atom methane molecules 
in the static field of zeolite ITQ-29.
We will assess the validity of our coarse-graining approach
by comparing the adsorption isotherms and occupancy distributions 
of the FG systems with their CG counterparts, and we
will draw conclusions in Section~\ref{sec:conclusions}.

\section{Local, Coarse-grained interactions \label{sec:CGpf}}

Our general FG model of reference will be a system of small 
guest molecules hosted inside an ordered microporous material, which is
represented as a network, $\mathcal{L}=\{\ell_1,\dots,\ell_M\}$, 
of $M$ pores
with \emph{local} connections,
meaning that the molecules inside a pore, e.g.~pore $i$,
interact with the inner surface of the pore itself, with the 
molecules inside the same pore, and with the molecules hosted 
in the $\nu$ neighboring pores.
Interactions with pores located beyond the first neighborhood are neglected
(this is often a fair assumption, since in several microporous materials, 
like LTA- and FAU-type zeolites, the pore size is 
approximately equal or larger than 12 \AA, which in most cases is near the customary cutoff 
radius for Lennard-Jones interactions).
The system is assumed to be in contact with a thermal bath and a reservoir of molecules,
so that both the temperature, $T$ 
(we will indicate with $\beta$ the `inverse temperature', $\beta=1/k_B T$, where $k_B$ 
is the Boltzmann's constant), and the chemical potential, $\mu$, are held fixed and 
uniform throughout the whole system, while the energy and the total number of guest molecules
are allowed to vary.

For every possible configuration of guest molecules in the system, we can
count how many of them fall within each pore, and then measure a
global occupancy configuration, $\{n_1,\dots,n_M\}$, indicating that
pore $1$ contains $n_1$ guests, pore $2$ contains $n_2$ of them, etc.
We assume then that
\begin{itemize}
\item[(i)] every single pore, say pore $i$,
contributes to the free energy
of the entire system by an amount $H_{n_i}$, and that
\item[(ii)] the interaction between two neighboring pores, 
  say $i$ and $j$, contribute by an additional amount $K_{n_i,n_j}$.
\end{itemize}
The quantities $Q_{n_i}$ and $Z_{n_i,n_j}$
can be conveniently introduced:
\begin{align}
  & Q_{n_i}= \exp\big(-\beta H_{n_i}\big),
  \label{eq:QQ}
  \\
  & Z_{n_i,n_j}= \exp\big(-\beta K_{n_i,n_j}\big)
  .
  \label{eq:HK}
\end{align}
$H$ and $Q$ are defined over properties of one single pore, 
therefore we will refer to either of them as `self-interaction terms'.
$K$ and $Z$ contain information about pore pairs,
and we will refer to either of them as `pair-interaction terms'.
The most
detailed description of the structure of the CG system is
provided by the global occupancy distribution,~\cite{Tunca2003}
${p}_\mu(n_1,\dots,n_M)$, i.e.~the probability of 
pore 1 having occupancy $n_1$, 
pore 2 having occupancy $n_2$, 
etc.,
\begin{align}
  {p}_\mu(n_1,\dots,n_M)
  =\frac{1}{\Xi_\mathrm{CG}}
  \prod_{i=1}^M e^{\beta\mu n_i} Q_{n_i}
  \prod_{j\in\mathcal{L}_i} \sqrt{Z_{n_i,n_j}}
  ,
  \label{eq:pGlobal}
\end{align}
where $\mathcal{L}_i$ is the list of the $\nu$ neighbors of pore $i$.
In Eq.~(\ref{eq:pGlobal}), the normalization constant 
$\Xi_\mathrm{CG}$ is the \emph{CG grand partition function}:
\begin{align}
  \Xi_\mathrm{CG}=
  \sum_{n_1}\cdots\sum_{n_M} 
  \prod_{i=1}^M e^{\beta\mu n_i} Q_{n_i}
  \prod_{j\in\mathcal{L}_i} \sqrt{Z_{n_i,n_j}}
  ,
  \label{eq:GCpfGlobal}
\end{align}
where the square root is introduced to correct for  
counting the pair-interaction terms twice.
The distribution in Eq.~(\ref{eq:pGlobal}) can be
easily sampled by Monte Carlo in the grand canonical
ensemble [see Supporting Information of our 
previous work~\cite{Pazzona2014}].

In Eq.~(\ref{eq:GCpfGlobal}), $Q_{n_i}$
plays the role of the `effective partition function of a single pore 
constrained to occupancy $n_i$'.
$Z_{n_i,n_j}$ instead plays the role of the `contribution to the
configuration integral of a pore pair constrained 
to occupancies $n_i,n_j$, due to the interaction of the
$n_i$ molecules in pore $i$ with the $n_j$ molecules in pore $j$'.

The scope of our coarse-graining approach here would be
to formulate CG interaction terms such that, once 
used in a CG (lattice) simulation,
they allow for the CG model to produce
a global occupancy distribution, ${p}_\mu(n_1,\dots,n_M)$,
in good agreement with its FG counterpart, $P_\mu(n_1,\dots,n_M)$
(throughout the whole paper, lowercase $p$'s will indicate 
CG probabilities, whereas capital $P$'s will refer to the 
FG system).
We used `would be' rather than `is' because, in practice, 
the $M$-variated histogram ${p}_\mu(n_1,\dots,n_M)$ 
can be estimated for none but the smallest systems.
Therefore, we will seek agreement in terms of
simpler (namely, uni- and bi-variated) distributions.
As long as the assumed locality of interactions holds,
we can reasonably expect that a good agreement in terms of 
local distributions will entail agreement also on a larger 
scale.

One important aspect we would like to remark is that
we want CG interactions to be \emph{local}, therefore
we require both $Q_{n_i}$ and $Z_{n_i,n_j}$ \emph{not} to 
depend on chemical potential, i.e.~we want the same set of 
self- and pair-interaction terms to be portable within 
a whole range of densities, from infinite dilution 
to saturation.

Let us now discuss the meaning of the 
interaction terms $Q_{n_i}$ and $Z_{n_i,n_j}$
on a statistical-mechanical basis.
$Q_{n_i}$ is commonly seen as the canonical partition function of
the pore $i$ when it contains exactly $n_i$ guest molecules,
i.e.~$Q_{n_i}=z_{n_i}/\Lambda^{3n_i}n_i!$ where
$\Lambda$ is the De Broglie thermal wavelength and
$z_{n_i}$ is the following configuration integral:
\begin{align}
  z_{n_i}= 
  \int\de\bv{r}_{i1} \cdots \int\de\bv{r}_{in_i}
  e^{-\beta U_i(\bv{r}^{}_{i1},\dots,\bv{r}^{}_{in_i})}
  ,
  \label{eq:SinglePorePFmolecular}
\end{align}
where $U_i$ denotes the potential energy experienced by 
the $n_i$ molecules hosted inside pore $i$, due to 
their interaction with the host material and with each other, 
given that their coordinates inside the pore are $\{\bv{r}_{i1},\dots,\bv{r}_{in_i}\}$.
In other words, the pore described by $Q_{n_i}$ is 
a small \emph{closed} system.
In principle, however, molecular configurations inside 
neighboring pores are correlated.
Therefore, assigning $Q_{n_i}$ a fixed value,
although being very convenient, might seem 
quite unnatural.
The pair term, $Z_{n_i,n_j}$ is thus introduced in order 
to account for such correlations.

The accepted meaning~\cite{Tunca2003} of $Z_{n_i,n_j}$
is that of the ratio between the configuration integral
of two pores with occupancies $n_i,n_j$ and the product of the
individual pore configuration integrals $z_{n_i}$ and $z_{n_j}$,
\begin{align}
  Z_{n_i,n_j} \sim &
  \frac{1}{z_{n_i}z_{n_j}}
  \int\de\bv{r}_{i1} \cdots \int\de\bv{r}_{in_i}
  \int\de\bv{r}_{j1} \cdots \int\de\bv{r}_{jn_i}
  \nonumber\\ & \times
  e^{-\beta U_{ij}(\bv{r}^{}_{i1},\dots,\bv{r}^{}_{in_i},\bv{r}^{}_{j1},\dots,\bv{r}^{}_{jn_j})}
  ,
  \label{eq:PorePairPFmolecular}
\end{align}
where $U_{ij}$
is the potential energy experienced by the molecules
inside pore $i$ and pore $j$ due to the interaction with the host material
and with each other, given that 
the $n_i$ molecules in pore $i$ are configured according to the 
coordinates $\{\bv{r}_{i1},\dots,\bv{r}_{in_i}\}$,
and that 
the $n_j$ molecules in pore $j$ are configured according to the 
coordinates $\{\bv{r}_{j1},\dots,\bv{r}_{jn_j}\}$.
With the symbol $\sim$ in (\ref{eq:PorePairPFmolecular}) 
we remark that we prefer to assume a weaker relation than
equality.
This is because relation~(\ref{eq:PorePairPFmolecular}) refers to a system made
of two pores, $i$ and $j$, respectively occupied by $n_i$ and $n_j$
guest molecules, as if it were `extracted' from the system where it belongs
and sampled separately from it, whereas in general the surroundings
of any pair of neighboring pores \emph{do} affect the correlations 
between them.

In a previous work~\cite{Pazzona2014} we proposed an 
estimation of effective free energies based on a very simple
reductionistic model, in which the surroundings
of a given pore were taken account of, but, in order
to derive an equation for the pair contributions
that could be solved straightforwardly, 
the neighbors' occupancies were all constrained to the same value.
In the next Section we will introduce a more accurate 
model in which the constraint on
the neighbors' occupancies is relaxed, and
mean-field (occupancy dependent) correction terms
are added to the free energy in the attempt to overcome 
the limitations of relation~(\ref{eq:PorePairPFmolecular}).

\section{Coarse-graining under the Interacting Pair Approximation \label{sec:IPA}}

\begin{figure*}[!t]
  \centering
  \includegraphics[width=0.99\textwidth]{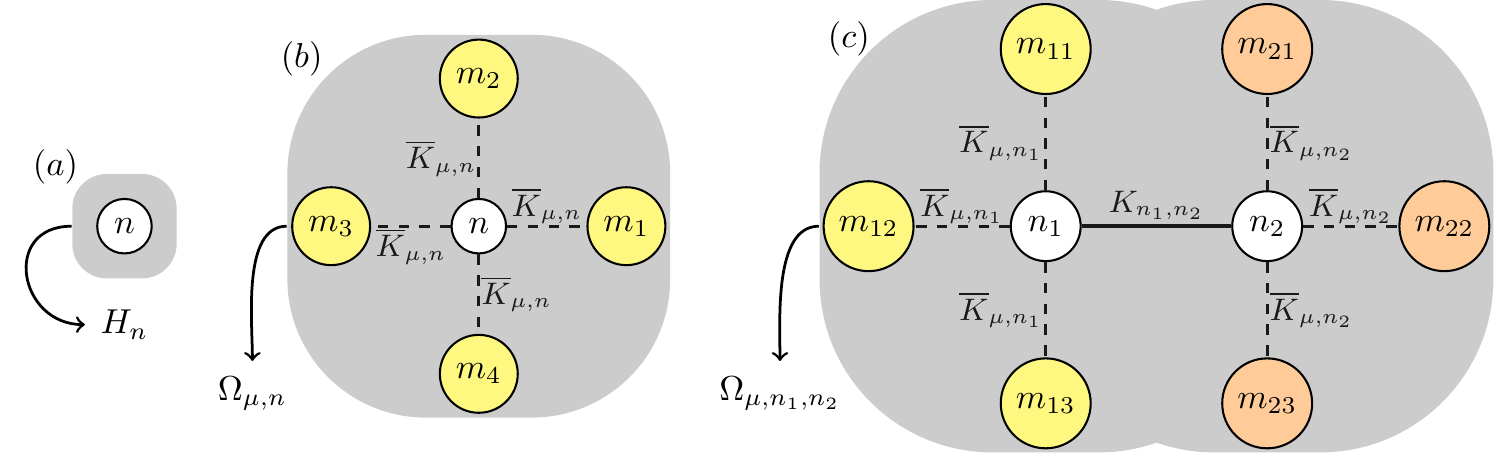}
  \caption{\footnotesize{A sketch of the role played by each interaction term 
      in the basic equations of our coarse-graining strategy. 
      In this graphic example, the reference FG system is a square lattice
      of pores, in which each pore is connected to $\nu=4$ neighbors.
      The pores represented by yellow circles are mean-field pores.
      In (a) we consider a single, $n$-occupied closed pore whose 
      equilibrium properties are related to the
      self-interaction term $H_n$.
      In (b) and (c) we consider the FG system as a whole, and from
      its equilibrium properties we derive the pair-interaction terms: 
      in (b) the $\nu$ neighbors of a single, $n$-occupied pore contribute to the
      CG potential, each one by adding a mean-field contribution
      $\overline{K}_{\mu,n}$ to the self-interaction $H_n$;
      in (c) the pores in a connected pair are assumed to interact with each 
      other through the non-mean-field pair term $K_{n_1,n_2}$ that 
      adds to the self-terms $H_{n_1}$ and $H_{n_2}$, 
      and their interactions with the rest of the system are approximated by
      two mean-field terms, $\overline{K}_{\mu,n_1}$ and 
      $\overline{K}_{\mu,n_2}$, each with multiplicity $\nu-1$.
  }}
  \label{fig:CGscheme}
\end{figure*}
Let us reformulate the problem in terms of simpler
probability mass functions than ${p}_\mu(n_1,\dots,n_M)$.
Temperature and volume will be assumed constant throughout the
entire discussion.
For a given value of chemical potential, $\mu$,
we will consider the following distributions:\\

{$p^o_\mu(n)$}:
  probability of a pore to be occupied by $n$ molecules,
  when interactions with \emph{all} the other pores are
  neglected;\\

{$p_\mu(n)$}: 
  probability of a pore to be occupied by $n$ molecules,
  with interactions with every one of the $\nu$ pore neighbors 
  represented as a mean-field, $\overline{K}_{\mu,n}$;\\

{$p_\mu(n_1,n_2)$}: 
  probability of a pore pair, made of pores 1 and 2, 
  to show the occupancy pair 
  $n_1,n_2$ with the effective interactions between
  the two pores given by $K_{n_1,n_2}$, the interactions
  between pore 1 and every one of its remaining $\nu-1$ neighbors
  represented as a mean-field $\overline{K}_{\mu,n_1}$, 
  and the interactions
  between pore 2 and every one of its remaining $\nu-1$ neighbors
  represented as a mean-field $\overline{K}_{\mu,n_2}$.\\

The distributions 
$p^o_\mu(n)$, $p_\mu(n)$, and $p_\mu(n_1,n_2)$
are defined in terms of the potential functions, which we call
\emph{CG potentials},
$\Omega^o_\mu(n)$, $\Omega_\mu(n)$, and $\Omega_\mu(n_1,n_2)$,
respectively, according to:
\begin{align}
  & p^o_\mu(n)= (\zeta^o_\mu)^{-1}\exp\{-\beta\Omega^o_{\mu}(n)\}
  ,
  \label{eq:PselfO}
  \\
  & p_\mu(n)= \zeta_\mu^{-1}\exp\{-\beta\Omega_{\mu}(n)\}
  ,
  \\
  & {p}_\mu(n_1,n_2)
  =
  \xi_\mu^{-1}
  \exp\{-\beta\Omega_{\mu}(n_1,n_2)\}
  ,
  \label{eq:pIPAConnection}
\end{align}
where $\zeta^o_\mu$, $\zeta_\mu$, and $\xi_\mu$ are normalization
constants, and, following the Bethe-Peierls mean-field 
approximation,~\cite{Bethe1935,Huang}
the CG potentials are defined as follows:
\begin{align}
  \Omega^o_{\mu}(n) = &
  -\mu n + H_{n}
  ,
  \label{eq:CGgp_single_closed}
  \\
  \Omega_{\mu}(n) = & \,\,
  \Omega^o_{\mu}(n)
  + \nu \overline{K}_{\mu,n}
  ,
  \label{eq:CGgp_single}
  \\
  \Omega_{\mu}(n_1,n_2) = & \,\,
  \Omega^o_{\mu}(n_1) + \Omega^o_{\mu}(n_2)
  + K_{n_1,n_2} \nonumber \\
  & + (\nu-1)(\overline{K}_{\mu,n_1}+\overline{K}_{\mu,n_2})
  .
  \label{eq:CGgp_pair}
\end{align}
$H_{n}$, the free energy of a \emph{closed} $n$-occupied pore,
and $K_{n_1,n_2}$, the contribution to the free energy
provided by the interaction between the $n_1$ molecules located in
pore $1$ and the $n_2$ molecules located in pore $2$, were already introduced
in Eq.~(\ref{eq:HK}).
By definition, $K_{n,0}=0$, i.e.~there is no 
effective interaction energy between
the molecules inside a pore and an empty pore.

Mean-field terms like $\overline{K}_{\mu,n}$,
are used as corrections to the free energy.
They can be thought as
$\overline{K}_{\mu,n}\sim\sum_m p_\mu(m|n) K_{n,m}$,
with $p_\mu(m|n)=p_\mu(n,m)/p_\mu(n)$,
even though, as we are going to show, there is 
no need to compute mean-field interactions explicitly.
In other words,
when we consider a single pore in the system, 
as in Eq.~(\ref{eq:CGgp_single}), 
$\overline{K}_{\mu,n}$ accounts for the interaction 
between the $n$ molecules inside that pore, 
and the molecules in its $\nu$ neighbors.
The number of such surrounding molecules,
although it is related to $\mu$,
is not specified anywhere,
therefore such $\nu$ neighbors can be thought
as mean-field pores.
When a pore pair of occupancy $(n_1,n_2)$ 
is considered instead, as we do in
Eq.~(\ref{eq:CGgp_pair}), we account for the rest of
the system in terms of $2(\nu-1)$
surrounding mean-field pores,
$\nu-1$ of which interact with cell 1 through
the potential $\overline{K}_{\mu,n_1}$, 
while the other $\nu-1$ ones
interact with cell 2 through the potential
$\overline{K}_{\mu,n_2}$.
In order to obtain a solvable system of equations,
we assume mean-field neighbors to not interact with 
each other.\\

The crucial point in Eqs.~(\ref{eq:CGgp_single})
and~(\ref{eq:CGgp_pair})
is that, although the mean-field terms
are $\mu$-dependent, the pair interaction terms, 
$K_{n_1,n_2}$, \emph{do not depend on $\mu$}.

In Fig.~\ref{fig:CGscheme} we sketched 
the role of the interaction terms used in 
Eqs.~(\ref{eq:CGgp_single_closed}), (\ref{eq:CGgp_single}), 
and~(\ref{eq:CGgp_pair}).
The closed-pore equation, Eq.~(\ref{eq:CGgp_single_closed}),
does not contain any mean-field term---in some sense, it is
`exact', meaning that if we were able to estimate with 
infinite accuracy the probability distribution $p^o_\mu(\cdot)$,
e.g.~by an infinitely long grand canonical sampling
[by the grand canonical Monte Carlo method (GCMC)~\cite{UMS}]
of a version 
of the FG system where only pore $1$ can be occupied and all the 
pores in the system stay empty,
we could retrieve $H_{n}$ from the difference $\Omega^o_{\mu}(n)-\Omega^o_{\mu}(n')$,
where $n'\neq n$,
knowing that $H_{0}=0$, or equivalently,
we could estimate $Q_{n}$ from the probability 
ratios $p^o_\mu(n)/p^o_\mu(n')$, knowing that $Q(0)=1$ can be used
as starting point:
\begin{align}
  \frac{
    Q_{n}
  }{
    Q_{n'}
  }
  =  e^{-\beta\mu(n-n')}
  \frac{
    p^o_\mu(n)
  }{
    p^o_\mu(n')
  }
  .
  \label{eq:Qself}
\end{align}
Resorting to ratios like $Q_{n}/Q_{n'}$ rather
than calculating every $Q_{n}$ directly from 
$p^o_\mu(n)=(\zeta^o_\mu)^{-1} e^{\beta\mu n}Q_n$
[see Eqs.~(\ref{eq:QQ}), (\ref{eq:PselfO}), 
and (\ref{eq:CGgp_single_closed})],
is motivated by the fact that we do not know 
in advance the normalization 
constant $\zeta^o_\mu$.

The ratio in Eq.~(\ref{eq:Qself}) does not depend on
chemical potential, meaning that, in principle, when
carrying out the calculation of the R.H.S.~of Eq.~(\ref{eq:Qself}),
one should recover the same result independently of the
value of $\mu$ at which the probabilities were evaluated.
In practice, however, numerical simulations are carried out
over a finite time. 
Therefore, when replacing $p^o_\mu(n)$ and $p^o_\mu(n')$
with $P^o_\mu(n)$ and $P^o_\mu(n')$, i.e.~the probabilities 
estimated from simulations of the FG system 
(with all the pores kept empty except for one), 
the R.H.S.~of Eq.~(\ref{eq:Qself}) will return
a slightly different value for each $\mu$, that is,
\begin{align}
  \frac{
    Q_{n}
  }{
    Q_{n'}
  }
  \approx  e^{-\beta\mu(n-n')}
  \frac{
    P^o_\mu(n)
  }{
    P^o_\mu(n')
  }
  .
  \label{eq:Qself_num}
\end{align}
A proper combination of the ratios in Eq.~(\ref{eq:Qself}) 
computed at different values of $\mu$ is the strategy we 
(successfully) used in our previous work~\cite{Pazzona2014} to 
obtain very reasonable results.

Once we computed the array of $Q$'s (or $H$'s) from  
GCMC on a single pore, we can proceed to
the evaluation of the pair-interaction parameters $K_{n_1,n_2}$
appearing in Eq.~(\ref{eq:CGgp_pair}).
By knowledge of the difference in CG potential 
\[
\Omega_{\mu}(n_1,n_2)-\Omega_{\mu}(n_1',n_2')
=
-\frac{1}{\beta} 
\ln\frac{
  {p}_{\mu}(n_1,n_2)
}{
  {p}_{\mu}(n_1',n_2')
}
,
\]
where $n_1'$ and $n_2'$ are chosen to be
not simultaneously equal to $n_1$ and $n_2$,
we can easily obtain an equation that relates them with 
$K_{n_1,n_2}-K_{n_1',n_2'}$
[or equivalently, with $Z_{n_1,n_2}/Z_{n_1',n_2'}$].
Eq.~(\ref{eq:CGgp_single}) can be used to eliminate the
mean-field terms,
and we obtain (for the sake of conciseness, we will express
the resulting equation in terms of the $Z_{n_1,n_2}$s):
\begin{align}
  \frac{
    Z_{n_1,n_2}
  }{
    Z_{n_1',n_2'}
  }
  = &
  \left(
  e^{-\beta\mu(n_1+n_2-n_1'-n_2')}
  \frac{
    Q_{n_1'} Q_{n_2'}
  }{
    Q_{n^{}_1} Q_{n^{}_2}    
  }
  \right)^\frac{1}{\nu}
  \nonumber\\ &\times
  \left(
  \frac{
    {p}_\mu(n_1') {p}_\mu(n_2')
  }{
    {p}_\mu(n_1) {p}_\mu(n_2)    
  }
  \right)^{1-\frac{1}{\nu}}
  \frac{
    {p}_\mu(n_1,n_2)
  }{
    {p}_\mu(n_1',n_2')
  }
  ,
  \label{eq:ZfracIPA}
\end{align}
with the corresponding free energy difference
given by Eq.~(\ref{eq:HK}).
In the R.H.S.~of Eq.~(\ref{eq:ZfracIPA}), 
the mean-field interactions, appearing in 
Eqs.~(\ref{eq:CGgp_single}) and~(\ref{eq:CGgp_pair}), 
are accounted for through 
the $1/\nu$ exponent on the first term 
(regarding the properties of a \emph{lone} cell), and through
the ratio involving single-cell probabilities, raised
to the power of $1-1/\nu$.

We can write down an equation by which 
the physical meaning of pair-interaction terms
will appear very intuitive.
To do so, 
we first introduce the observed-to-expected (o/e) ratio,
$C_\mu(n_1,n_2)={p}_\mu(n_1,n_2)/{p}_\mu(n_1){p}_\mu(n_2)$,
whose deviation from unity is a measure of the
correlations between the neighbor pore occupancies $n_1,n_2$,
and the ratio $D_\mu(n)={p}_\mu(n)/p^o_\mu(n)$ which measures
the amount by which the mean-field neighborhood of a single
pore causes its properties to deviate from the 
closed-pore case.
Now, if we consider that the guest-guest interaction between 
two pores with no guests inside is null 
(so that $Z_{0,0}=1\Rightarrow K_{0,0}=0$),
then we can see that the pair terms have the following meaning:
\begin{align}
  K_{n_1,n_2} = &
  -\frac{1}{\beta}
  \Big[
  \ln C_\mu(n_1,n_2)
  + 
  \frac{1}{\nu}
  \ln\big[D_\mu(n_1) D_\mu(n_2)\big]
  \nonumber\\
  & -
  \ln C_\mu(0,0)
  - 
  \frac{2}{\nu}
  \ln\big[D_\mu(0)\big]
  \Big]
  ,
  \label{eq:KpairFEorder_pre}
\end{align}
where the terms $\ln C_\mu(0,0)$ and 
$\frac{2}{\nu}\ln D_\mu(0)$ are related to the occupancy
pair $0,0$, taken as a reference state.
All terms in the R.H.S.~of Eq.~(\ref{eq:KpairFEorder_pre})
depend on $\mu$, but for each $\mu$ they change such as
to return the same value.
According to Eqs.~(\ref{eq:CGgp_single_closed}),
(\ref{eq:CGgp_single}), and~(\ref{eq:CGgp_pair}),
for a given pair of neighboring occupancies $n_1,n_2$,
the R.H.S.~of Eq.~(\ref{eq:ZfracIPA}) must be the same 
at all chemical potentials.
Therefore, one can formally remove the dependence on $\mu$ from
Eq.~(\ref{eq:KpairFEorder_pre}),
by integrating it over a range 
that goes from $\mu_\mathrm{i}$, corresponding to very low density, to
$\mu_\mathrm{f}$, corresponding to very high density, close to saturation.
In this way, the terms related to 
the reference state, i.e.~the ones in which both the pores of the pair 
are empty, will appear as a single constant:
\begin{align}
  K_{n_1,n_2} = &
  -\frac{1}{\big(\mu_\mathrm{f}-\mu_\mathrm{i}\big)\beta}
  \int_{\mu_\mathrm{i}}^{\mu_\mathrm{f}}
  \mathrm{d}\mu
  \Big[
  \ln C_\mu(n_1,n_2)
  \nonumber\\ &
  + 
  \frac{1}{\nu}
  \ln\big[D_\mu(n_1) D_\mu(n_2)\big]
  \Big]
  + \mathrm{const}.
  \label{eq:KpairFEorder}
\end{align}
Although only formally, Eq.~(\ref{eq:KpairFEorder}) provides us
with the meaning of the CG pair interaction terms, consistent
with the assumptions made in Eqs.~(\ref{eq:CGgp_single}) and~(\ref{eq:CGgp_pair}),
that is,
\emph{except for a constant term, 
contributions to the pair free energy $K_{n_1,n_2}$ come from
the correlation between the neighbor occupancies $n_1$ and $n_2$,
and from the effect of the local surroundings on each of the two 
pores (divided by the pore connectivity $\nu$),
at \emph{all} the chemical potentials in the range
$\mu_\mathrm{i}<\mu<\mu_\mathrm{f}$}.

As it is, Eqs.~(\ref{eq:ZfracIPA}) and~(\ref{eq:KpairFEorder})
cannot be used directly for the calculation of the
pair-interaction terms, because they require
knowledge of the coarse-grained ${p}_\mu$ distributions, 
which are unknown.
Therefore, we need a key assumption in order to convert
our mean-field formulation of this problem into an 
operative coarse-graining strategy. 
Our proposal is to replace the unknown distribution
${p}_\mu$, with the 
distribution obtained by numerical simulation of the 
FG system, $P_\mu$.
This amounts to saying that,
at any $\mu$ in the range
$\mu_\mathrm{i}<\mu<\mu_\mathrm{f}$,
the approximation
\begin{align}
  P_\mu(n_1,n_2)\approx{p}_\mu(n_1,n_2),
  \label{eq:IPA}
\end{align}
holds for every occupancy pair $n_1,n_2$.
We will refer to 
the approximation~(\ref{eq:IPA}), together with 
Eqs.~(\ref{eq:CGgp_single}) and~(\ref{eq:CGgp_pair}),
as \emph{Interacting Pair Approximation} (IPA), to emphasize
that we considered the pair of pores as a physical region 
that is not kept away from the rest of the system, 
but rather interacts with its surroundings through 
mean-field correction terms.
As an immediate consequence of the fact that relation~(\ref{eq:IPA})
is an approximation, once we replaced the theoretical $p_\mu$ with 
the numerical distribution $P_\mu$, we have that the
R.H.S.~of Eq.~(\ref{eq:ZfracIPA}) becomes only approximately
equal to the ratio $Z_{n_1,n_2}/Z_{n_1',n_2'}$:
\begin{align}
  \frac{
    Z_{n_1,n_2}
  }{
    Z_{n_1',n_2'}
  }
  \approx &
  \left(
  e^{-\beta\mu(n_1+n_2-n_1'-n_2')}
  \frac{
    Q_{n_1'} Q_{n_2'}
  }{
    Q_{n^{}_1} Q_{n^{}_2}    
  }
  \right)^\frac{1}{\nu}
  \nonumber\\ &\times
  \left(
  \frac{
    {P}_\mu(n_1') {P}_\mu(n_2')
  }{
    {P}_\mu(n_1) {P}_\mu(n_2)    
  }
  \right)^{1-\frac{1}{\nu}}
  \frac{
    {P}_\mu(n_1,n_2)
  }{
    {P}_\mu(n_1',n_2')
  }
  .
  \label{eq:ZfracIPA_num}
\end{align}
In other words, \emph{in practice}, different chemical potentials
will contribute differently to the estimation of the ratio
$Z_{n_1,n_2}/Z_{n_1',n_2'}$.
Among all such contributions, we can identify some 
values of $\mu$ that we want to contribute 
more than other ones, because they correspond to
situations in which
the pore occupancies $n_1$, $n_2$, $n_1'$, and $n_2'$
are visited frequently enough for us to reckon
that our estimation of the probabilities 
${p}_\mu(n_1)$, ${p}_\mu(n_2)$, 
${p}_\mu(n_1,n_2)$,
${p}_\mu(n_1')$, ${p}_\mu(n_2')$,
and ${p}_\mu(n_1',n_2')$
is accurate enough (e.g., if the probabilities are
larger than some threshold).
Conversely, we want $\mu$ values at which those pore 
occupancies are sampled rarely to contribute \emph{less},
since in those cases our estimation of the probabilities
is expected to be rather inaccurate.
Extreme situations, i.e.~values of $\mu$ at which some or all 
of the occupancies $n_1$, $n_2$, $n_1'$, and $n_2'$ 
are never sampled, should then give no contribution to 
$Z_{n_1,n_2}/Z_{n_1',n_2'}$.
This might cause some $Z_{n_1,n_2}$ to
remain unknown,~\cite{Pazzona2014} but this does not really represent an issue,
as long as the computable entries of the matrix $Z$ 
ensure that the probability distribution that
can be obtained by simulation of the resulting coarse-grained 
system and their FG counterparts reasonably match at all chemical 
potentials.
Further details are discussed in the Supplementary Material,
along with the description of two possible routes 
for the estimation of the interaction terms
$Q_n$ and $Z_{n_1,n_2}$---
in the first one, 
reported also in our previous work,~\cite{Pazzona2014} 
and indicated here as `one-chemical-potential-at-a-time' 
(OCT), in a first stage we make use of 
Eqs.~(\ref{eq:Qself}) and~(\ref{eq:ZfracIPA})
recursively \emph{for each chemical potential},
thus obtaining $\mu$-dependent
CG interactions, and in a second stage we remove 
the $\mu$-dependency through a weighted average.
In the second one, that we indicate as 
`choose-the-best-ratio' (CBR) we select
the $\mu$ for which
the R.H.S.~of Eq.~(\ref{eq:ZfracIPA_num}) can be regarded
as the best representative of the ratio
$Z_{n_1,n_2}/Z_{n_1',n_2'}$, e.g.~by using, as selection
criterion, how large and how similar the probabilities
${P}_\mu(n_1,n_2)$ and ${P}_\mu(n_1',n_2')$
are, and then, we use the ratios we selected to calculate 
recursively the individual entries of the matrix $Z$.
Essentially, the differences in the interaction matrix $Z$
obtained using either of the two methods are very small,
while a much more crucial role is played by the accuracy
in the probability histograms evaluation from GCMC.

\section{Comparison with previous models \label{sec:compare}}
It is worthwhile to compare our coarse-graining (IPA) approach,
with the more drastic assumption in which a pair of neighboring
pores is treated as if it was uncorrelated with the rest of 
the FG system.~\cite{TuncaFord2002,Tunca2003,Tunca2004}
We will indicate the latter assumption as
\emph{Non-Interacting Pair Approximation}
(NIPA).

NIPA relies on relation~(\ref{eq:PorePairPFmolecular}) 
\emph{taken as if it were an equality}.
To compare IPA and NIPA, we find it convenient 
to write the IPA equation for the pair interaction terms, 
i.e.~Eq.~(\ref{eq:ZfracIPA}), as follows:
\begin{align}
  \frac{
    Z_{n_1,n_2}
  }{
    Z_{n_1',n_2'}
  }
  = &
  \left(
  \frac{
    p^o_\mu(n_1') p^o_\mu(n_2')
  }{
    p^o_\mu(n_1) p^o_\mu(n_2)    
  }
  \right)^\frac{1}{\nu}
  \left(
  \frac{
    {p}_\mu(n_1') {p}_\mu(n_2')
  }{
    {p}_\mu(n_1) {p}_\mu(n_2)    
  }
  \right)^{1-\frac{1}{\nu}}
  \nonumber\\ &\times
  \frac{
    {p}_\mu(n_1,n_2)
  }{
    {p}_\mu(n_1',n_2')
  }
  ,
  \label{eq:ZfracIPA2}
\end{align}
If relation~(\ref{eq:PorePairPFmolecular}) was an
equality, we could drop the
mean-field terms in Eq.~(\ref{eq:CGgp_pair}),
thus obtaining the NIPA equation for the pair interactions:
\begin{align}
  \frac{
    Z^*_{n_1,n_2}
  }{
    Z^*_{n_1',n_2'}
  }
  = &
  \frac{
    p^o_\mu(n_1') p^o_\mu(n_2')
  }{
    p^o_\mu(n_1) p^o_\mu(n_2)    
  }
  \,
  \frac{
    {p}^*_\mu(n_1,n_2)
  }{
    {p}^*_\mu(n_1',n_2')
  }
  ,
  \label{eq:ZfracNIPA}
\end{align}
where ${p}^*_\mu(n_1,n_2)$ is the probability of a pair
of neighboring pores \emph{separated from the rest of the system}
to show the occupancy pair $n_1,n_2$, given that the
chemical potential is $\mu$.
The first major problem with NIPA is that
the adsorption isotherm of a closed pair is,
at high densities, different from the 
adsorption isotherm of the FG system as a whole
(as shown in the Supplementary Material 
for the case of the Lennard-Jones system we will 
discuss in Section~\ref{sec:LJ}).
Therefore, in general, the NIPA and IPA 
occupancy distributions are expected to be also different.
Moreover,
we can see by comparing the NIPA Eq.~(\ref{eq:ZfracNIPA})
with the IPA Eq.~(\ref{eq:ZfracIPA2}), that,
when switching from NIPA to IPA,
inclusion of the mean-field corrections causes the
single-pore NIPA term in the R.H.S.~of Eq.~(\ref{eq:ZfracNIPA}),
\[
\frac{
  p^o_\mu(n_1') p^o_\mu(n_2')
}{
  p^o_\mu(n_1) p^o_\mu(n_2)    
},
\]
to split into two factors, 
in the R.H.S.~of Eq.~(\ref{eq:ZfracIPA2}),
\[
  \left(
  \frac{
    p^o_\mu(n_1') p^o_\mu(n_2')
  }{
    p^o_\mu(n_1) p^o_\mu(n_2)    
  }
  \right)^\frac{1}{\nu}
  \left(
  \frac{
    {p}_\mu(n_1') {p}_\mu(n_2')
  }{
    {p}_\mu(n_1) {p}_\mu(n_2)    
  }
  \right)^{1-\frac{1}{\nu}}
  ,
\]
that is, one independent-pore contribution, raised to the power
of $1/\nu$, where a single pore of occupancy $n$ is taken 
as if it were a closed system, and one correlated-pore contribution, raised
to the power of $1-1/\nu$ (and therefore, more important than the first one), 
which istead relates the properties of a single pore to its 
surroundings in the FG system, \emph{via} mean field 
correction terms.
Therefore, use of the NIPA matrix $Z^*$ will in general ensure 
the correct coarse-graining of only a special version of the FG system, in which only 
two pores are non-empty, but not of the FG system as a whole.
Since the correlations between any pore and its surroundings becomes
of crucial importance at high density, the IPA matrix $Z$ is expected
to provide, in general, a more accurate CG representation.

\begin{figure}[!t]
  \centering
  \includegraphics[width=0.49\textwidth]{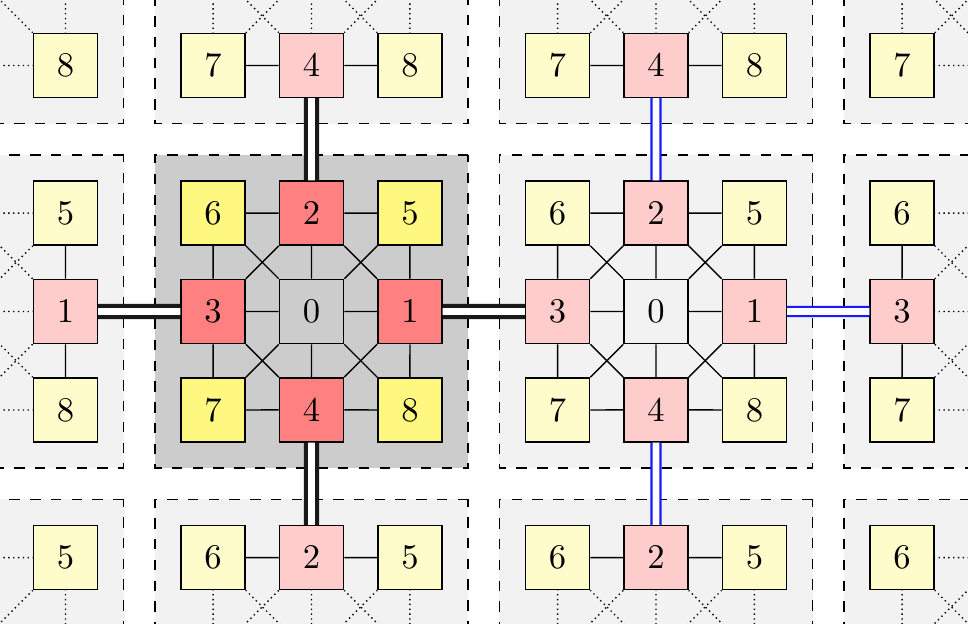}
  \caption{\footnotesize{Structure of the lattice-gas we studied
      in this work to compare the IPA with the NIPA coarse-graining
      approach.
      Sites, which can assume either state 0 (empty) or 1 
      (singly occupied) are represented as small squares, 
      which can be grouped into cells (gray shades).
      A number is assigned to each site within every cell 
      to distinguish from one another.
      Site-site interactions are pairwise, and they take place between
      connected sites---connections are displayed as lines, which are
      thin if the connected pair entirely belongs to one cell, and
      thicker (and doubled) if they connect two sites that belong to different 
      cells.
  }}
  \label{fig:Lattice}
\end{figure}
Before we proceed further with the next Section,
it is worth mentioning the conditions under which
the IPA strategy described here reduces to the coarse-graining
strategy we proposed in a previous work,~\cite{Pazzona2014}
where a CG equation for the ratio
$Z_{n_1,n_2}/Z_{n_1-1,n_2}$ was derived by constraining the
occupancies in the neighborhood of a given pore to the same
value.
By letting $n_1'=n_1-1$ and $n_2'=n_2$, we can rewrite 
Eq.~(\ref{eq:ZfracIPA}) as
\begin{align}
  \frac{
    Z_{n_1,n_2}
  }{
    Z_{n_1-1,n_2}
  }
  = &
  \left(
  e^{-\beta\mu}
  \frac{
    Q_{n_1^{}-1}
  }{
    Q_{n^{}_1}
  }
  \right)^\frac{1}{\nu}
  \left(
  \frac{
    {p}_\mu(n_2|n_1)
  }{
    {p}_\mu(n_2|n_1-1)
  }
  \right)^{1-\frac{1}{\nu}}
  ,
  \label{eq:ZfracIPAbis}
\end{align}
where ${p}_\mu(n_2|n_1)$ is the conditional probability of a pore,
belonging to a pair of neighboring pores,
to have occupancy $n_2$, given that the other pore has occupancy 
$n_1$.
We can see that the basic CG expression 
we proposed in our previous work is retrieved when
the last factor in the R.H.S.~of 
Eq.~(\ref{eq:ZfracIPAbis}) can be neglected (i.e.~when it is $\sim 1$).
This happens under the approximation 
${p}_\mu(n_2|n_1)\approx{p}_\mu(n_2|n_1\pm 1)$,
that represents a less general case where
the conditional distribution ${p}_\mu(\cdot|n_1)$ does not
vary much when the neighbor occupancy $n_1$ is \emph{slightly} varied,
thus implying weak (even though still non-null) lateral 
correlations.

\section{Simulations and discussion \label{sec:numsim}}

In this Section we apply both the IPA and the NIPA approaches
to a lattice-gas system of interacting boolean sites 
(Section~\ref{sec:lattice}) and to a Lennard-Jones 
system of confined particles (Section~\ref{sec:LJ}).
All the simulations were carried out by standard Metropolis 
GCMC.~\cite{UMS}

\subsection{Lattice-gas with repulsive interactions \label{sec:lattice}}
The local contributions given by the $Q$ array 
and the NIPA pair-interaction matrix $Z^*$ 
can be (numerically) calculated exactly for a lattice-gas where cells 
are comprised of a small number of $n_\mathrm{max}$ mutually exclusive sites,
since in that case the integrals in Eqs.~(\ref{eq:SinglePorePFmolecular}) 
and~(\ref{eq:PorePairPFmolecular}) reduce to summations over 
a large but finite number of configurations.
This makes lattice-gases an invaluable tool for comparing different
coarse-graining strategies, such as IPA and NIPA.

Our lattice-gas here is a square lattice of cells, each one
made of nine sites arranged as a square as well.
Every site can be either empty (occupancy 0) or occupied by one particle 
(occupancy 1).
Neighboring sites, say $i$ and $j$, interact with each other 
(lateral interactions) repulsively, according to the  
interaction energy of $\epsilon$.
With the aim of increasing the correlations,
in some simulations we `extended' the FG interactions by
including an attractive interaction parameter, $\psi$, 
based on the number of occupied neighbors around each site:
\begin{align}
  E(\bv{s})= \sum_{\langle i,j \rangle} s_i s_j 
  \Big[ \epsilon + \psi(M_i) + \psi(M_j) \Big]
  ,
  \label{eq:FGhamil}
\end{align}
where the sum runs over all the pairs of neighboring sites,
and
$s_i$ and $s_j$ are the occupancies of sites $i$ and $j$,
according to the occupancy configuration $\bv{s}$
of the whole FG lattice.
$M_i$ and $M_j$ are defined as the total
occupancy in the neighborhood, respectively, of
site $i$ (including the occupancy of $j$) and of
site $j$ (including the occupancy of $i$), and
\begin{align}
  \psi(M)= 
  \left\{
  \begin{array}{cc}
    \phi, & M \geq M_0 \\
    0, & M<M_0 
  \end{array}
  \right.
  ,
  \label{eq:FGext}
\end{align}
where $\phi<0$.
The energy $\psi(M)$ adds to the interaction between 
two neighboring sites if the number of occupied 
neighbors of each of them becomes equal or larger 
than some threshold value $M_0$, which we set at $M_0=4$.

In Fig.~\ref{fig:Lattice}, the structure of a portion of
the lattice is depicted.
Interacting sites are joined by 
lines, that are either thin or thick, respectively 
in the case of intra-cell and inter-cell connections.
Intercell connections are represented in Fig.~\ref{fig:Lattice} 
as `double' connections, but this does not imply that
the interaction energy is doubled.
\begin{figure}[t!]
  \centering
  \includegraphics[width=0.4\textwidth]{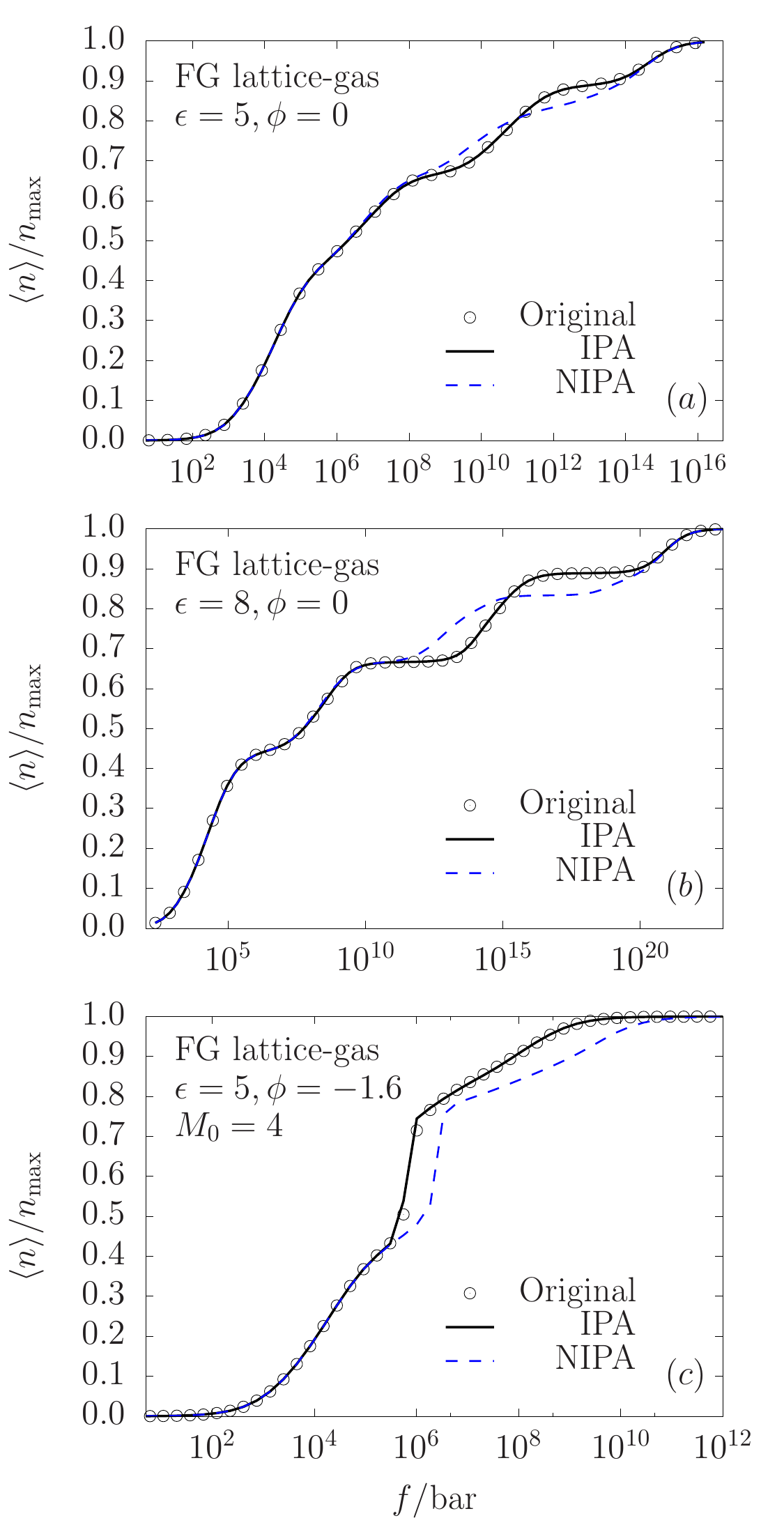}
  \caption{\footnotesize{Adsorption isotherms 
      for the lattice-gas system under different
      interaction setups.
      In (a) and (b), the site-site interaction is purely 
      repulsive (it amounts respectively to 4 and 8 kJ mol$^{-1}$).
      In (c), the lateral interaction is set at 4 kJ mol$^{-1}$,
      but extended attractive interactions are added.
      Results for the FG system are depicted as empty
      circles, whereas solid black lines are used for
      IPA and dashed blue lines for NIPA results.
      For the sake of readability, 
      we reduced
      the density of points in the FG scatter plot
      to one half of the actual dataset.
  }}
  \label{fig:L1L2}
\end{figure}
GCMC simulations of this FG system under different 
setups of the interaction parameters 
were performed
at several values of chemical potential,
chosen such as to ensure that the resolution was at least
of two density points between each interval
$(\langle n\rangle,\langle n\rangle+1)$ in the average cell 
occupancy.
In Fig.~\ref{fig:L1L2} we show
results for the following parameter settings: 
(a) $\epsilon=4$ kJ mol$^{-1}$ and $\phi=0$,
(b) $\epsilon=8$ kJ mol$^{-1}$ and $\phi=0$, and
(c) $\epsilon=4$ kJ mol$^{-1}$, $\phi=-1.6$ kJ mol$^{-1}$, and $M_0=4$.
For every chemical potential, two simulations were performed.
In the first one, inter-cell interactions were neglected
and the $Q$ terms were evaluated from~(\ref{eq:Qself_num}).
In the second simulation, we included inter-cell interactions
and evaluated the $Z$ interaction terms through~(\ref{eq:ZfracIPA_num}).
Every simulation was carried out
over a number of steps that varied from $N=10^6$ to $10^7$ moves, 
equally (and randomly) distributed among displacement, 
insertion, and deletion attempts.
Simulations of both IPA and NIPA CG systems were performed
through GCMC as well, but over a smaller number of steps
($N\sim 10^5$) due to the much faster convergence to equilibrium.
The results reported in this work are for lattice systems
of $4\times 4$ cells.
Larger systems were explored ($6\times 6$ and $8\times 8$)
for a smaller number of GCMC moves and of
chemical potential values, and gave results that were
indistinguishable from the ones obtained for the
$4\times 4$ cases.

\begin{figure*}[t!]
  \centering
  \includegraphics[width=0.99\textwidth]{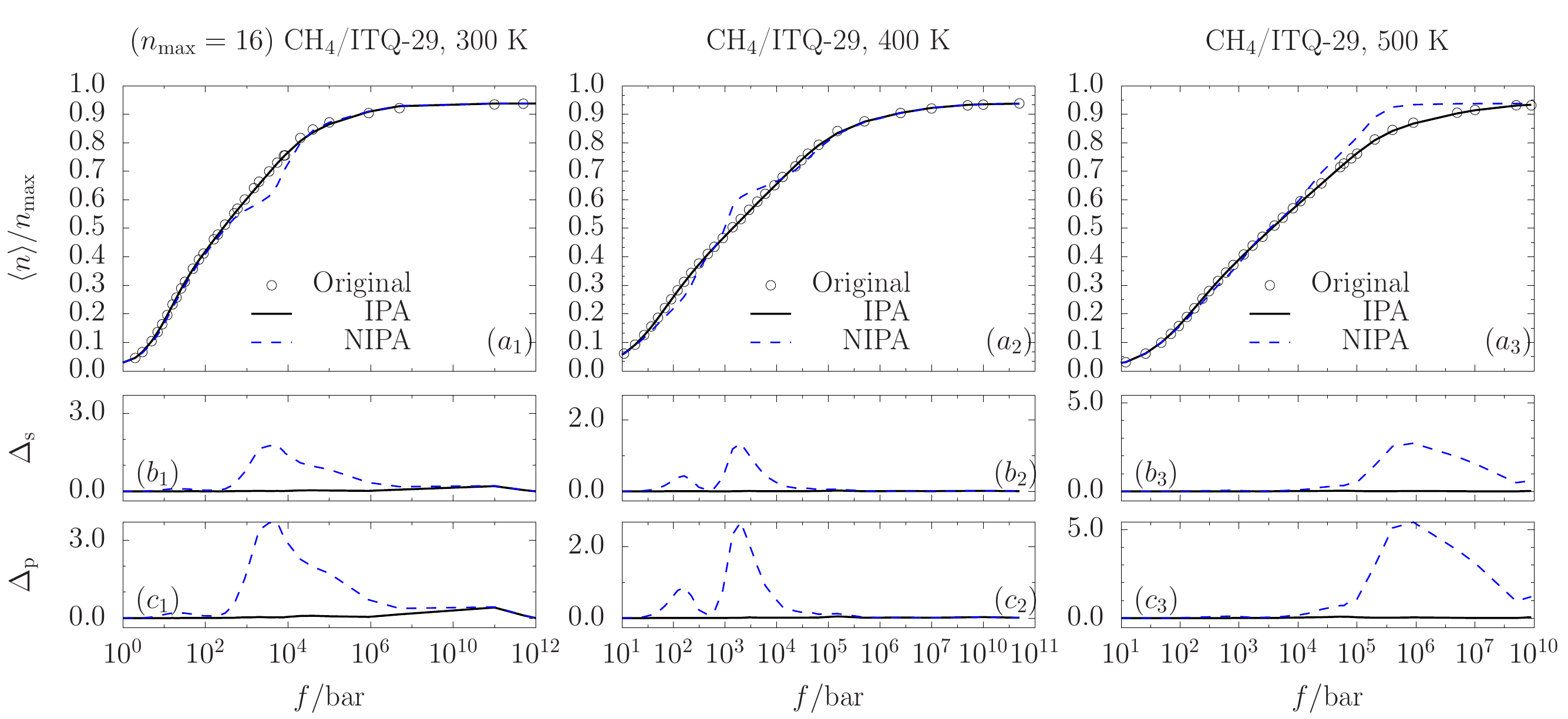}
  \caption{\footnotesize{
      Adsorption isotherms and Kullback-Leibler divergences for 
      a system of a Lennard-Jones methane molecules (united atom
      approximation) under the static field of zeolite ITQ-29
      at the temperatures 300, 400, and 500 K.
      In subfigures $a_1$, $a_2$, and $a_3$, adsorption isotherms are
      shown (empty circles: FG lattice system;  black solid lines: IPA;  
      blue dashed lines: NIPA).
      In subfigures $b_1$, $b_2$, and $b_3$, the Kullback-Leibler
      divergence for the occupancy distribution 
      of one single cell are shown [see Eq.~(\ref{eq:KullSelf})].
      Subfigures $c_1$, $c_2$, and $c_3$, refer instead to the occupancy 
      distribution one pair of neighboring cells 
      [see Eq.~(\ref{eq:KullPair})].
      Black solid lines represent divergences between FG and IPA systems,
      blue dashed lines represent divergence between FG and NIPA systems.
  }}
  \label{fig:CH4_in_ITQ29_300_500}
\end{figure*}
For both the IPA and the NIPA coarse-graining, the results
we reported were
obtained through the CBR approach described in 
Section~\ref{sec:IPA}.
However, both OCT and CBR provided nearly the same results.
Adsorption isotherms, i.e.~plots of the density (expressed as the 
average cell occupancy, $\langle n\rangle$, divided by the
total number of sites per cell, $n_\mathrm{max}$) 
\emph{vs.}~the fugacity (here meant as $f=f_0 e^{\beta\mu}$, where $f_0=1$ bar),
are reported in Fig.~\ref{fig:L1L2}, and they show that the
NIPA approach starts failing at intermediate-high
densities, where intercell correlations become important.
On the other hand, IPA provides isotherms
(see Fig.~\ref{fig:L1L2}) and occupancy distributions
(see Supplementary Material) 
in good agreement with the FG system at all densities.
In particular, in the example shown in Fig.~\ref{fig:L1L2}a,
at high densities, pair correlations are non-negligibly affected 
by the presence of the other neighbors of both cells of the pair, 
and this causes the adsorption isotherm of the whole FG system to exhibit
curvature changes that are not well reproduced by NIPA.
In Fig.~\ref{fig:L1L2}b, a more repulsive site-site 
interaction enhances this phenomenon, and
the isotherm tends toward a step-like shape as 
repulsion is increased.
In this case, the more quantitative agreement provided by the IPA
approach is even more evident.
The isotherm in Fig.~\ref{fig:L1L2}c is related to a
more extreme case, where,
due to the increasingly important effect of the 
attractive contribution from $\psi(M)$ to the total energy, 
see Eqs.~(\ref{eq:FGhamil}) and ~(\ref{eq:FGext}),
site correlations extend to the second neighborhood.
One can immediately figure out that extended interactions
may cause cell pairs to be correlated very differently, 
depending on whether we consider 
every pair as if it was part of a larger portion of the system 
(as in the IPA approach), or
as if it evolved on its own, detached from the rest of
the system (as is the NIPA approach).
As a consequence of the balance between repulsive
and attractive interactions, a larger step appears in
the isotherm at intermediate densities,
and as the density approaches the step 
(for $\langle n\rangle/n_\mathrm{max}$ between 0.4 and 0.5),
the NIPA method fails.
On the contrary, IPA better preserves the shape 
of the original system, indicating that, also in this case
the cell-cell correlations induced by more complicated FG
interactions are well represented through the inclusion of
the mean-field terms in Eqs.~(\ref{eq:CGgp_single}) 
and~(\ref{eq:CGgp_pair}).

We remark that in the calculations above, the NIPA interaction terms
were evaluated as exact sums rather than through
simulations of a pair of cells, so they are not affected by any
accuracy issue, whereas the IPA
interaction terms were calculated straight from the distributions 
obtained from simulations of the FG system---therefore, 
contrarily to the NIPA case, IPA parameters are supposed 
to be not immune to noise and accuracy issues (related to the fact 
that low-probability occupancies are unavoidably sampled less frequently,
and then less accurately, than the high-probability ones);
despite everything, the IPA reveals the most accurate of the two.
However, as we will see in Section~\ref{sec:LJ},
in systems where the structure is determined by a much smoother
potential energy function, the difference between IPA and NIPA, although
undeniably present, appears less marked and starts becoming non-negligible
at higher densities.

\subsection{Lennard-Jones particles under the influence of an external field \label{sec:LJ}}
Methane molecules, represented by the united atom approximation 
as Lennard-Jones (LJ) spheres, confined in the all-silica zeolite ITQ-29 
(formerly called ZK4) have been widely used in the literature as a host-guest
system to test statistical-mechanical theories, 
adsorption-diffusion models,
methods for the calculation of free energy profiles, 
and coarse-graining approaches under various
computational environments (like kinetic Monte Carlo and 
Cellular Automata).~\cite{Demontis1997_JCPB,TuncaFord1999,TuncaFord2002,Tunca2003,Tunca2004,Demontis2005,Smit2004,Smit2005,Smit2006,Demontis2008JPCB,Pintus2011,Pintus2015}
The ITQ-29 framework is particularly interesting
because of its peculiar structure
of relatively wide pores (when compared to methane size), 
called $\alpha$-cages ($\sim 11.4$ \AA~in diameter), 
arranged in a
simple cubic network ($\nu=6$), 
and interconnected through narrower eight-ringed 
windows ($\sim 4.5$ \AA~in diameter), allowing the passage of one
methane molecule at a time.
We modeled guest-guest and host-guest interactions according to the 
force fields used by Dubbeldam \emph{et al.}~\cite{Smit2005}
with a cutoff of $12$ \AA,
and, since the zeolite flexibility does not affect significantly 
the sorption properties of methane (although it would be not negligible
for larger molecules~\cite{Vlugt2002}), a pre-tabulation of the 
host-guest potential energy on a grid of $\sim 0.2$ \AA~of spacing
allowed for a significant reduction of the CPU time of the 
simulations.~\cite{June1990}
Our framework system consisted of a grid of $4\times 4\times 4$ pores,
corresponding to $2\times 2\times 2$ unit cells (the ITQ-29 unit cell 
we used consisted of eight pores).
GCMC simulations were carried out using the standard Metropolis 
acceptance-rejection method for displacements, insertions, and
deletions.~\cite{UMS}
Such MC moves where performed in equal proportions, within 
a total number of post-equilibration steps that varied
from $\sim 10^6 N_\mathrm{uc}$ to $\sim 10^8 N_\mathrm{uc}$,
with $N_\mathrm{uc}$ as the average number of molecules per unit cell.
The temperatures we investigated were 100, 200, 300, 400, and 500 K.
The fugacities were chosen in such a way as to explore loadings 
more or less uniformly (at least two points within each loading 
interval from $\langle n\rangle$ to $\langle n\rangle+1$)
from $\sim 0.1$ up to $\sim 14.5$ molecules per pore.
In all cases, methane molecules were not allowed to enter
the sodalite cages nor the double six-ringed cages.
At~100 K, 
due to the very low acceptances at the highest loadings,
simulations were carried out up to $\sim 12$ molecules per pore.
\begin{figure}[t!]
  \centering
  \includegraphics[width=0.4\textwidth]{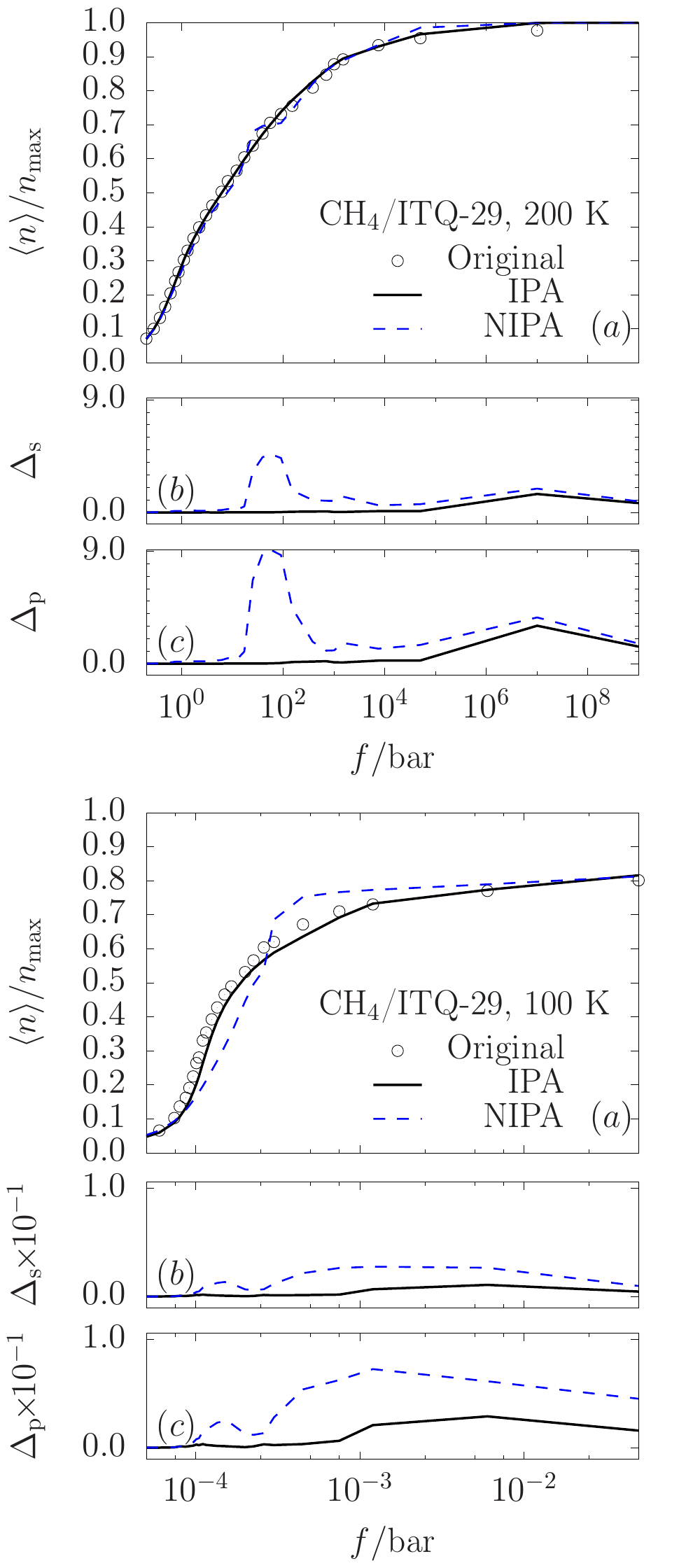}
  \caption{\footnotesize{
      Adsorption isotherms and Kullback-Leibler divergences for 
      a system of a Lennard-Jones methane molecules (united atom
      approximation) under the static field of zeolite ITQ-29
      at the temperatures 100, and 200 K.
      The meaning of dots, line types, and line colors, is the same
      as in Fig.~\ref{fig:CH4_in_ITQ29_300_500}.
  }}
  \label{fig:CH4_in_ITQ29_LowT}
\end{figure}
Due to the very simple (cubic) topology of the pore network, and 
since methane-methane interactions across non-first neighboring
pores can be safely neglected,~\cite{Pazzona2014}
the CH$_4$/ITQ-29 system is especially suited for testing the 
IPA coarse-graining scheme as well.

In Fig.~\ref{fig:CH4_in_ITQ29_300_500} we compare results for the
temperatures 300, 400, and 500 K.
At such temperatures, 
the CBR approach provided slighlty better IPA representations,
whereas slightly better NIPA results were obtained
by using the OCT protocol.
Besides adsorption isotherms, we wanted to give the reader a
quick idea on how the use of IPA rather than NIPA
affects the occupancy distributions of the
CG model, in comparison with the distributions
that emerge from the GCMC simulations of the FG system.
Since two kinds of histogram were constructed out of GCMC simulations
at every chemical potential
(one univariate histogram for the probability of any pore to have occupancy $n$,
and one bivariate histogram for the probability of any pore pair to show the occupancy
pair $n_1,n_2$), in order to be able to visualize the results on a single
figure per system, here we decided to compare occupancy distributions
through the Kullback-Leibler (KL) divergence, that we used according to
the symmetric definition given by Kullback and Leibler in their
original article.~\cite{Kullback1951}
We will refer to $\Delta_\mathrm{s}$ as the KL divergence
for the probability distribution of a single pore,
and to $\Delta_\mathrm{p}$ as the KL divergence for the probability
distribution of a pore pair:
\begin{align}
  & \Delta_\mathrm{s}= 
  \sum_{n} \Big(
    P_\mu(n) -p^\mathrm{cg}_\mu(n) \Big) 
    \ln \frac{P_\mu(n)}{p^\mathrm{cg}_\mu(n)}
  ,
  \label{eq:KullSelf}
  \\
  & \Delta_\mathrm{p}=
  \sum_{n_1}\sum_{n_2} \Big(
  P_\mu(n_1,n_2) - p^\mathrm{cg}_\mu(n_1,n_2) \Big)
\ln \frac{P_\mu(n_1,n_2)}{p^\mathrm{cg}_\mu(n_1,n_2)}
  ,
  \label{eq:KullPair}
\end{align}
where $P_\mu$ and $p^\mathrm{cg}_\mu$ refer respectively to the occupancy
distribution of the FG system and of one of two possible CG systems
(IPA and NIPA).
Based on the resulting FG distributions, 
we set the maximum pore occupancy at $n_\mathrm{max}=15$.
We included more detailed comparisons of the occupancy distributions
in the Supplementary Material.

As we anticipated at the end of Section~\ref{sec:lattice},
the discrepancies between IPA and NIPA are less
evident here than in the case of lattice-gases with repulsive interactions,
due to the smoothness of the LJ potentials.
Nevertheless, the IPA approach shows to be the most accurate
in all the cases reported, proving its robustness despite its
simplicity.
At low loadings, both approaches provide a reasonable
agreement between CG and FG systems, but at intermediate-high
loadings,
non-negligible KL divergences between the NIPA and the 
FG distributions appear, in correspondence with discrepancies 
in the adsorption isotherms (as expected), and they are 
much more pronounced than the ones we find for the IPA case.
We believe this is due to the presence of the mean-field
terms in the basic equations of the IPA approach,
Eqs.~(\ref{eq:CGgp_single}) and~(\ref{eq:CGgp_pair}), which
satisfactorily accounts for the effect of the whole neighborhood
of each pore.

In Fig.~\ref{fig:CH4_in_ITQ29_LowT} we report results at lower
temperatures, namely, 200 and 100 K.
The IPA parameters that produced the CG plots in Fig.~\ref{fig:CH4_in_ITQ29_LowT}
were calculated by the CBR method at 200 K,
and by the OCT method at 100 K, whereas
the NIPA parameters were evaluated through the OCT method
at both temperatues.
Also in these cases, the difference between the parameters obtained 
by the two methods is not so much evident, and we made
our choice based on slight discrepancies.

At these temperatures, correlations between neighboring
pores become more evident.
Noticeably, at 200 K, while the IPA and NIPA isotherms are 
approximately in the same (good) agreement with the FG system,
the occupancy distributions are not, and the IPA results
are closer to the FG distributions, especially at low densities.

At the temperature of 100 K the NIPA approach fails to provide
a reasonable agreement even at low loadings, indicating that
in this case the occupancy of each pore is
seriously affected by the occupancies
\emph{in the whole neighborhood}.
Including only one
neighbor in the statistical description of CG interactions, 
as NIPA prescribes,
does not allow the CG model to reproduce, not even 
partially, the correlations observed in the FG 
system.
The FG occupancy distributions at 100 K become highly non-central
for all but the lowest loadings (this can be seen very clearly 
in the figure reported in the Supplementary Material), 
and we noticed that, although still resulting more satisfactory 
than NIPA, the agreement in the CG occupancy distributions as 
provided by the IPA approach becomes less striking than at higher 
temperatures.
In particular, bimodality, that we also observed for the system at
200 K, and that correspond to states with two coexisting 
phases,~\cite{GuemezPhysA01,Cheung1993,Muller2006} is not accurately reproduced.
We believe this not to be an issue of the mean-field corrections as
they are formulated in Eqs.~(\ref{eq:CGgp_single}) and~(\ref{eq:CGgp_pair}),
but rather a limitation of the pairwise nature of the CG 
potential model.
Inclusion of other correction terms that depend on collective,
but still local, variables, may 
further improve the agreement in situations where
correlations between every pore and all its
neighbors are \emph{very} large.~\cite{Wagner2017}
This will be the subject of forthcoming investigations.\\

\section{Conclusions \label{sec:conclusions}}
We investigated the coarse-graining of host-guest 
systems of small molecules adsorbed in a regular 
porous material, described in terms of
occupancy distributions rather than fine-grained
configurations of molecular positions.
In such a reductionistic representation, the 
interaction field is based on the free energy
of every single pore, defined as a function of
its occupancy (i.e.~the number of molecules it
hosts), plus effective contributions to the 
free energy coming from the interactions between
neighboring pore pairs.
By means of a very simple system, i.e.~a lattice-gas 
where local free energies can be calculated
exactly, we have shown that the currently accepted approximation 
in which the pair interaction is assumed to be the same 
whether the pore pair is kept within the full fine-grained 
system it belongs, or it is made independent of 
its surroundings~\cite{TuncaFord1999,TuncaFord2002,Tunca2003,Tunca2004} 
(we referred to it as NIPA, \emph{non-interacting pair approximation}),
turns out to be inaccurate at high densities, where 
the interactions between every pore pair and its 
neighborhood induce stronger correlations.
In Lennard-Jones systems, where interactions are much
smoother than in lattice-gases, the inadequacy of the NIPA
approach is slightly less evident but, 
apart from the case of high temperatures 
(around room temperature and above)
and low sorbate density, still leads to 
non-negligible discrepancies between the fine-grained 
system and its coarse-grained counterpart.
We improved the calculation of coarse-grained 
interactions by establishing a relation
between local occupancy distributions
of the fine-grained systems and the properties of
a coarse-grained, occupancy-based model, 
that we called IPA 
(\emph{interacting pair approximation}), where
the effect of the surroundings on both single
pores and pore pairs is taken account of
\emph{via} mean-field terms.
As a result, the pore pair interactions appear as if
they were entirely related to the local pore-pore correlations,
and to the discrepancy between the properties of a
closed single pore and those of a pore which instead
does interact with its neighbors.
We remark that,
although in the basic IPA equations, mean-field corrections
depend on chemical potential (i.e.~they are density-dependent),
the resulting coarse-grained interactions do \emph{not} depend
on it, i.e., their local nature is preserved.
We presented results for the coarse-graining of lattice-gases
with repulsive interactions, and for a host-guest model
of methane molecules (treated as Lennard-Jones spheres) 
confined in zeolite ITQ-29.
In every case we studied, 
the IPA approach provided noticeably better results than NIPA.
In the majority of cases, the 
the agreement between the properties of the coarse-grained systems obtained
under the IPA approach, and the properties of the original, fine-grained system,
was excellent.

\bibliographystyle{aip}
\bibliography{bibliografia2.8}

\end{document}